\documentclass[preprint,showpacs,preprintnumbers,amsmath,amssymb]{revtex4}

\usepackage{graphicx,subfigure}
\usepackage{dcolumn}
\usepackage{bm}
\usepackage{amsmath}

\newcommand{\be}{\begin{equation}}
\newcommand{\ee}{\end{equation}}
\newcommand{\bea}{\begin{eqnarray}}
\newcommand{\eea}{\end{eqnarray}}

\newcommand{\gev}{\rm GeV}

\newcommand{\rme}{\rm e}
\newcommand{\rmi}{\rm i}
\newcommand{\rmd}{\rm d}

\begin{document}

\bibliographystyle{apsrev}

\preprint{UAB-FT-615}

\title{Unified Model for Inflation and Dark Matter}

\author{Gabriel Zsembinszki}

\affiliation{Grup de F{\'\i}sica Te{\`o}rica and Institut de
F{\'\i}sica d'Altes
Energies\\Universitat Aut{\`o}noma de Barcelona\\
08193 Bellaterra, Barcelona, Spain}


\date{\today}

\begin{abstract}
We present a model which contains a new complex scalar field
$\Psi$, interacting with a new real scalar field, $\chi$, and an
effective potential whose symmetry is almost exact, being
explicitly broken by Planck-scale effects. We study the
possibility of relating {\em inflation} with the {\em dark matter}
of the universe. We find that for exponentially small explicit
breaking, the model accounts for a period of inflation in the
early universe and gives a dark matter candidate particle.
\end{abstract}

\maketitle

\section{Introduction}\label{Introduction}
Cosmology has made in the last few years enormous progress,
especially in the accuracy of observational data, which is taken
using technologies more and more precise. While in the past not
too remote it seemed almost inconceivable, nowadays we may even
talk about "precise cosmology". This is why, in analogy with the
Standard Model (SM) of particle physics, many physicists are
already talking about a "standard model" of cosmology. As
suggested by recent observations of type Ia supernovas (SNIa)
\cite{Supernova}, the matter power spectrum of large scale
structure (LSS) \cite{LSS} and anisotropy of the cosmic microwave
background radiation (CMB) \cite{WMAP3}, the universe is presently
dominated by too types of mysterious fluids: dark energy (DE),
which has negative pressure whose consequence is to accelerate the
expansion of the universe, and dark matter (DM), which is
non-relativistic non-baryonic matter, very weakly coupled to
normal matter and that only has gravitational effects on it.

The most simple explanation for DE is a cosmological constant
$\Lambda$, but it raises another problem because its expected
value is many orders of magnitude larger than the value suggested
by observations. Another possible explanation is the existence of
a slowly rolling scalar field, called quintessence, which is
displaced from the minimum of its potential and started to
dominate the energy density of the universe recently.

The same observations indicate that the universe is isotropic and
homogeneous at large scales and spatially flat, for which in the
old cosmological picture there is no reasonable explanation. The
most successful and simple solution to the flatness and
homogeneity problems is given by inflation \cite{Inflation}, which
in its simplest version is defined as a short period of
accelerated expansion of the early universe caused by a single
dominating scalar field, the inflaton. In addition, inflation
gives the most popular mechanism of generation of cosmological
fluctuations, which were the seed for the structure formation in
our universe.

Although the SM based on the gauge group $SU(3)\times SU(2) \times
U(1)$ is a solid theoretical construction able to accommodate all
existing empirical data, it leaves many deep questions unanswered
when trying to explain the origin and nature of the new
ingredients introduced by modern cosmology, such as, for example,
the inflaton, the DE and the DM. Thus, there are reasons to
believe that the SM is not the ultimate theory and one has to look
for extensions of it. If we are able to discover a theory that
indeed goes beyond the SM, it will probably contain new
symmetries, either local, or global.

A lot of effort has been done in studying global symmetries at
high energies \cite{quantum_coherence}, especially in trying to
clarify the issue of quantum coherence loss in the presence of
wormholes. It was argued that the loss of coherence opens the
possibility that currents associated with global symmetries are
not exactly conserved. Even if incoherence is not observed in the
presence of wormholes, it was argued that other interesting
consequences may emerge, such as the appearance of operators that
violate global symmetries, of arbitrary dimensions, induced by
baby universe interactions. There are other reasons to expect that
quantum gravity effects break global symmetries: global charges
can be absorbed by black holes which may evaporate, "virtual black
holes" may form and evaporate in the presence of a global charge,
etc.

In this context, the authors of \cite{Kallosh:1995hi} argue that
if global symmetries are broken by virtual black holes or topology
changing effects, they have to be exponentially suppressed. In
particular, in order to save the axion theory, the suppression
factor should have an extremely small value $g<10^{-82}$. This
suppression can be obtained in string theory, if the stringy mass
scale is somewhat lower than the Planck-scale, $M_{\rm str}\sim
2\times 10^{18} \gev$. Thus we expect to have an exponential
suppression of the explicit breaking of global symmetries.

Even with such an extremely small explicit breaking, one can see
that very interesting consequences may appear. In \cite{paper1}
(from now on Paper 1), it was shown that, when a global symmetry
is spontaneously broken in the presence of a small explicit
breaking, the resulting pseudo-Golstone boson (PGB) can be a DM
particle. In \cite{paper2} (from now on Paper 2), a similar study
was made, but the purpose was to show that the resulting PGB could
be a quintessence field explaining the present acceleration of the
universe. In addition, based on the idea forwarded by Frieman and
Rosenfeld \cite{Rosenfeld:2005mt}, the model of Paper 2 also
incorporated inflation. In this way, the two periods of
accelerated expansion may have a common origin.

There is previous work related to explicit breaking of global
symmetries \cite{Ross} and to Planck-scale breaking
\cite{Lusignoli}. Cosmological consequences of some classes of
PGBs are discussed in \cite{Hill}.

Here, we extend the model in Paper 1 to also include inflation.
The way we do it is similar to the work in Paper 2, the difference
being that here we want the resulting PGB to be a DM particle, in
contrast with Paper 2 where it was a quintessence field. Our
result is that the parameter of the explicit breaking should be
exponentially suppressed, $g< 10^{-30}$, as in Paper 1, but the
level of suppressions is not that high as in the case of Paper 2,
where a much smaller $g\sim 10^{-119}$ was needed in order for the
PGB to explain DE. Inflation may occur here at scales as low as
$V^{1/4}\sim 10^{10}\gev$.

The paper is structured as follows: in section \ref{Model} we make
a short presentation of the model and then focus on the main
features of it: inflation and dark matter. In section
\ref{Results} we present our numerical results and, finally, in
section \ref{Conclusions} we make a discussion and give the
conclusions. Technical details are given in the Appendix
\ref{Appendix}.

\section{The model}\label{Model}
The model is, basically, the same as in Paper 2, so we just recall
it here shortly. It contains a new complex scalar field, $\Psi$,
charged under a certain global $U(1)$ symmetry, interacting with a
massive real scalar field, $\chi$, neutral under $U(1)$. It also
contains a $U(1)-$symmetric potential
\begin{equation}
V_{\rm sym}(\Psi,\chi) =\frac14\lambda(|\Psi|^2-v^2)^2
+\frac12m_{\chi}^2\chi^2+\left(\Lambda^2-
\frac{\kappa^2|\Psi|^2\chi^2}{4\Lambda^2}\right)^2 \label{Vsym1}
\end{equation}
where $\lambda$ and $\kappa$ are coupling constants, $m_{\chi}$
and $\Lambda$ are some energy scales and $v$ is the $U(1)$
spontaneous symmetry breaking (SSB) scale.

The interaction term in (\ref{Vsym1}) is of inverted hybrid type
\cite{Ovrut:1984qp, Lyth:1996kt} and can be realized in
supersymmetry using a globally supersymmetric scalar potential
\cite{Lyth:1996kt}. However, in the present paper we are not
preoccupied about the underlying theory in which this model can be
realized, instead we only study the phenomenology of the potential
(\ref{Vsym1}).

Next, we allow terms in the potential that {\it explicitly} break
$U(1)$. These terms are supposed to come from physics at the
Planck-scale, and without knowledge of the exact theory at that
scale, we introduce the most simple effective $U(1)-$breaking term
\cite{nonsym_terms}
\begin{equation}
 V_{\rm non-sym}(\Psi)=-g\frac1{M_{\rm P}^{n-3}}|\Psi|^n\left(\Psi
\rme^{-\rmi\delta}+\Psi^{\star}
\rme^{\rmi\delta}\right)\label{Vnon_sym1}
\end{equation}
where $g$ is an effective coupling, $M_{\rm P}\equiv G_{\rm
N}^{-1/2}$ is the Planck-mass and $n$ is an integer ($n>3$).

Summarizing, our effective potential is
\begin{equation}
V(\Psi,\chi) = V_{\rm sym}(\Psi,\chi) + V_{\rm
non-sym}(\Psi)-C\label{V_tot}
\end{equation}
where $C$ is a constant that sets the minimum of the effective
potential to zero.

By writing the field $\Psi$ as
\begin{equation}
\Psi =\phi\, {\rme}^{\rmi\tilde\theta}\label{Psi}
\end{equation}
we envisage a model in which the radial field $\phi$ is the
inflaton, while the angular field $\tilde\theta$ is associated
with a DM particle.

Thus, in this paper we consider the possibility of having a
unified model of inflation and DM, improving in this way the model
presented in Paper 1. We also want to present a more detailed
numerical analysis of the part regarding inflation, which could
also apply to the inflationary model of Paper 2.

\subsection{Inflation}\label{Inflation}
We first revisit the conditions that should be accomplished by our
model in order to correctly describe the inflationary period of
expansion of the universe. Inflation is supposed to have occurred
in the early universe, when the energies it contained were huge.
Thus, the appropriate term to deal with when describing inflation
is the symmetric term $V_{\rm sym}$, while the non-symmetric one
can be safely neglected, being many orders of magnitude smaller
than $V_{\rm sym}$. Taking into account (\ref{Psi}), we may write
\begin{equation}
V_{\rm sym}(\phi,\chi)=\Lambda^4+\frac12 M^2_{\chi}(\phi)\chi^2+
\frac{\kappa^4\phi^4\chi^4}{16\Lambda^4}
+\frac14{\lambda}(\phi^2-v^2)^2-C \label{Vsym2}
\end{equation}
where $M^2_{\chi}(\phi)\equiv m_{\chi}^2 -\kappa^2\phi^2$, and we
have also included the constant $C$. As commented above, $\phi$ is
the inflaton field and $\chi$ is an auxiliary field that assists
$\phi$ to inflate.

We assume that initially the fields $\phi$ and $\chi$ are in the
vicinity of the origin of the potential, $\phi=\chi=0$. At that
point, the first derivatives of the potential are zero in both
$\phi-$ and $\chi-$directions, but the second derivatives have
opposite signs:
 \be
    \left.\frac{\partial^2V_{\rm sym}(\phi,\chi)}{\partial
    \phi^2}\right|_{\phi,\chi=0} =-\lambda v^2<0 \ee \be
    \left.\frac{\partial^2V_{\rm sym}(\phi,\chi)}{\partial
    \chi^2}\right|_{\phi,\chi=0} =m_{\chi}^2
    >0.
 \ee

This means that $\chi$ remains trapped at the false minimum in
$\chi-$direction of the potential, $\chi=0$, while $\phi$ becomes
unstable and can roll down in the direction given by $\chi=0$. If
the potential in $\phi-$direction is sufficiently flat, $\phi$ can
have a slow-roll and produce inflation. This regime lasts until
the curvature in $\chi-$direction changes sign and inflation has a
sudden end through the instability of $\chi$, which triggers a
waterfall regime and both fields rapidly evolve towards the
absolute minimum of the potential. The critical point where
inflation ends is given by the condition
\begin{equation}
M_{\chi}^2(\phi)=m_{\chi}^2-\kappa^2\phi^2=0 \label{Mass_of_chi}
\end{equation}
so that during inflation $\phi<\phi_{\rm c}
=\frac{m_{\chi}}{\kappa}\lesssim v$.

The constraints related to the inflationary aspects of the model
are the same of Paper 2. Let us just summarize them here.

\begin{itemize}
 \item vacuum energy of field $\chi$ should dominate: $\frac14\lambda v^4\ll\Lambda^4$
 \item small $\phi-$mass as compared to $\chi-$mass: $|m_{\phi}^2|
 =\lambda v^2\ll m_{\chi}^2\lesssim \kappa^2 v^2$
 \item slow-roll conditions: $\epsilon\equiv \frac1{16\pi}\left(
 \frac{V'}V\right)^2\ll 1$, $|\eta|\equiv \left|\frac1{8\pi}\frac{V''}V
 \right|\ll 1$
 \item rapid variation of $M_{\chi}^2(\phi)$ at the critical point:
 $|\Delta M_{\chi}^2(\phi_{\rm c})|>H^2$
 \item fast roll of $\phi$ after $\chi$ gets to its minimum:
 large $\left.(\partial V_{\rm sym}/\partial\phi)\right|_{\chi_{\rm min}}$
\end{itemize}

These conditions have to be satisfied in order for the hybrid
inflationary mechanism to work. There are other constraints
related to fairly precise observational data:

\begin{itemize}
  \item
   sufficient number of e-folds of inflation $N(\phi)
 =(8\pi)/M_{\rm P}^2\int_{\phi_{\rm end}}^{\phi}
(V_{\rm sym}/V'_{\rm sym})\rmd\phi$ in order to solve the flatness
and the horizon problems. The required number depends on the
inflationary scale and on the reheating temperature, and is
usually comprised between 35 for low-scale inflation and 60 for
GUT-scale inflation

\item
   the amplitude of the primordial curvature power-spectrum
   produced by quantum fluctuations of the inflaton field should
   fit the observational data \cite{WMAP3}, ${\cal P_R}^{1/2}\simeq
   4.86\times 10^{-5}$

\item
   the spectral index $n_{\rm s}$ should have the right value
   suggested by observations of the CMB \cite{WMAP3},
   $n_{\rm s}=0.951^{+0.015}_{-0.019}$ (provided tensor-to-scalar ratio
   $r\ll 1$).
\end{itemize}

Combining all the above constraints we obtain the following final
relations that should be satisfied by the parameters of our model:
$\lambda\ll\kappa^2$ and $v<M_{\rm P}$. We also obtain the
dependence of some of the model parameters on the SSB scale $v$
(for more details, see Paper 2 and the Appendix \ref{Appendix}).
These will be used in section \ref{Results} for a numerical study.
The range of values of the scale $v$ will be fixed by the
requirement that $\theta$ is a DM candidate.

\subsection{Dark matter}\label{dark_matter}

As stated above, our idea is that $\theta$, the PGB that appears
after the SSB of $U(1)$ in the presence of a small explicit
breaking, can play the role of a DM particle. Thus, after the end
of inflation, $\theta$ finds itself in a potential given by the
term $V_{\rm non-sym}$
\begin{equation}
V_{\rm non-sym}(\phi,\theta)=
-2\,g\frac{\phi^{n+1}}{M_P^{n-3}}\cos{\tilde\theta}
\end{equation}
where (\ref{Psi}) has been used in (\ref{Vnon_sym1}) and the
change of variables $\tilde\theta\longrightarrow\tilde\theta
+\delta$ has been made. In Paper 1 it was shown that for
exponentially small $g$ the evolutions of the two components of
$\Psi$ are completely separated, so that we expect
$\theta-$oscillations to start long after $\phi$ has settled down
at its vacuum expectation value (vev),
\begin{equation}
\langle\phi\rangle\simeq M_P^{1/3}v^{2/3}.\label{vev_phi}
\end{equation}

A detailed study of the cosmology of the $\theta-$particle was
made in Paper 1, for the lowest possible value $n=4$. We do not
want to enter into details here, but just to make use of the
results of that work to obtain the values of the parameters of our
model. The only difference here is the fact that the vev of the
radial field $\phi$ is different from $v$, so that the constraints
obtained in Paper 1 on $v$ will apply here on
$\langle\phi\rangle$. This will also affect the angular field
$\tilde\theta$, which is here normalized as $\tilde\theta \equiv
\theta/\langle\phi\rangle$.

Due to the small explicit breaking of the $U(1)$ symmetry,
$\theta$ acquires a mass which depends on both $g$ and
$\langle\phi\rangle$
\begin{equation}
m_{\theta}^2=2g\left(\frac{\langle\phi\rangle}{M_P}\right)^{3}M_P^2
\label{thetamass}
\end{equation}
and this is why we should find constraints on both
$\langle\phi\rangle$ and $g$ in order for $\theta$ to be a
suitable DM candidate. The constraints that should be imposed come
from various astrophysical and cosmological considerations:

\begin{itemize}
\item
  $\theta$ should be a stable particle, with lifetime $\tau_{\theta}>t_0$, where
  $t_0$ is the universe's lifetime
\item
  its present density should be comparable to the present DM density
  $\Omega_{\theta}\sim\Omega_{DM}\sim 0.25$
\item
  it should not allow for too much energy loss and rapid cooling of
  stars \cite{Raffelt_book}
\item
  although stable, $\theta$ may be decaying at present, and
  its decay products should not distort the diffuse photon background
\end{itemize}
In Paper 1, all these constraints have been studied in detail.
Because $\theta$ is massive, it can decay into two photons or two
fermions, depending on its mass. The lifetime of $\theta$ depends
on the effective coupling to the two photons/fermions and on its
mass, which in turn depends on the two parameters
$\langle\phi\rangle$ and $g$. It was shown that for the
interesting value of $\langle\phi\rangle$ and $g$ for which
$\theta$ can be DM, the resulting $\theta-$mass has to be
$m_{\theta}<20$ eV, so the only decaying channel is into two
photons.

\begin{figure}[htb]
\includegraphics[width=11cm, height=7.5cm]{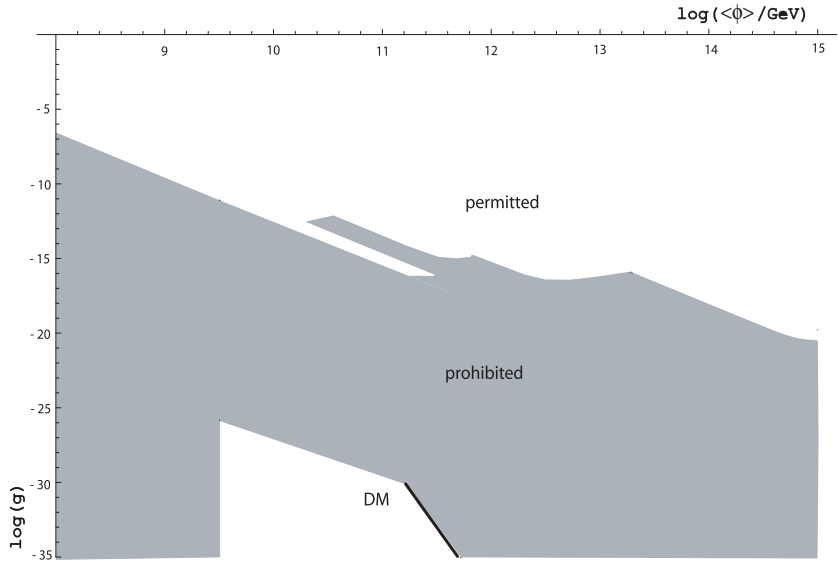}
\caption{Permitted and prohibited regions in the
$(\langle\phi\rangle,g)-$plane, taken from Paper 1. The
interesting points are those which are near the line labelled
"DM". }\label{fig1}
\end{figure}

There are different mechanism by which $\theta$ particles can be
produced, as explained in Paper 1: $(a)$ thermal production in the
hot plasma, and $(b)$ non-thermal production by $\theta-$field
oscillations and by the decay of cosmic strings produced in the
SSB. All these may contribute to the present energy density of
$\theta$ particles, which was computed in Paper 1. By requiring it
to be comparable to the present DM energy density of the universe,
we obtain a curve in the space of parameters $\langle\phi\rangle$
and $g$, illustrated in Fig.\ref{fig1} as the line labelled "DM".

In Paper 1 it was argued that there are some similarities between
our $\theta$ particle and the QCD axion \cite{axion}. This is why
when investigating its production in stars, we can apply similar
constraints. The strongest one comes from the fact that $\theta$
may be produced in stars and constitute a novel energy loss
channel, and these considerations put a limit on
$\langle\phi\rangle$, but not on $g$
\begin{equation}
\langle\phi\rangle> 3.3\times 10^9\, {\rm GeV}. \label{limvnucl}
\end{equation}

Another aspect about $\theta$ is that, although stable, a small
fraction of its population may be decaying today and the resulting
photons may produce distortions of the diffuse photon background.
Thus, the calculated photon flux coming from $\theta-$decay is
constraint to be less than the observed flux (see Paper 1 for
details).

By combining astrophysical and cosmological constraints, we can
obtain the interesting values for $\langle\phi\rangle$ and $g$ for
which $\theta$ is stable and its density is comparable to the DM
density. These values can be easily read from Fig.\ref{fig1},
being situated along the line labelled "DM". Note that there is an
upper limit on $g$, which corresponds to a lower limit on $\langle
\phi\rangle$, i.e.
 \begin{equation}
   g< 10^{-30}, \;\; \langle\phi\rangle >10^{11}\, {\rm GeV}.
   \label{g_cond2}
 \end{equation}
These also put a lower limit on the value of $U(1)$ breaking
scale,
 \be
    v\sim\left(\frac{\langle\phi\rangle ^3}{M_{\rm P}}\right)^{1/2}
    > 10^{7}\,\gev.\label{v_value}
 \ee

Finally, according to the study made in Paper 1, for value
$\langle \phi\rangle<7.2\times 10^{12}$ GeV, $\theta$ particles
can be produced both thermally and non-thermally, but in the
region characterized by $\langle\phi\rangle>10^{11}$ GeV, the
dominant energy density corresponds to non-thermally produced
$\theta$ particles. Moreover, for $\langle\phi\rangle>7.2\times
10^{12}$ GeV, only non-thermal production is possible.

\section{Numerical results}\label{Results}

The constraints enumerated in subsection \ref{Inflation} will
determine the values of some of the model parameters. Our scope is
to make a general analysis of how these parameters depend on $v$.
For $\lambda$ we obtain an exact formula (see Appendix
\ref{Appendix})
 \be
   \lambda =4.4\times
   10^{-12}\frac{\phi_0^2(v^2-\phi_0^2)^2}{(v^2-3\phi_0^2)^3}
 \ee
where $\phi_0$ is the value of the inflaton field when the scale
$k=0.002\ {\rm Mpc}^{-1}$ crossed the horizon during inflation and
its $v-$dependence can be obtained numerically, see the Appendix
\ref{Appendix}.

The exact formulae for the other parameters are too complicate to
be shown here. Nevertheless, things get simplified in the limit
$\lambda v^4\ll \Lambda^4(\equiv v\ll M_{\rm P})$, for which we
obtain
  \be \Lambda
    \simeq 1.6\times
    10^{-3}\left[\frac{\phi_0^2(v^2-\phi_0^2)^2}{(v^2-3\phi_0^2)^2}
    \right]^{1/4}M_{\rm P}^{1/2}\sim \lambda^{1/4}v^{1/2}
    M_{\rm P}^{1/2}
  \ee
and
  \be
    C\simeq \frac34\Lambda^4\left(\lambda v^4/\Lambda^4\right)^{1/3}
    \ll \Lambda^4.
  \ee
We notice that $C$ can be neglected, as compared to $\Lambda^4$,
in this limit.

The coupling $\kappa$ is only constrained by the condition
$\lambda\ll \kappa^2$, so that it can have any arbitrary value
satisfying this inequality. In our numerical study we took the
value $\kappa=10^{-2}$. The mass of $\chi$, namely $m_{\chi}$, can
have any arbitrary value satisfying $m_{\chi}< \kappa v$, but for
the sake of simplicity we set it to $m_{\chi}=\kappa v/2$, without
loss of generality.

\begin{figure}[]
\begin{picture}(60,500)(50,50)
\put
(-150,408){\includegraphics[width=7.5cm,height=5cm]{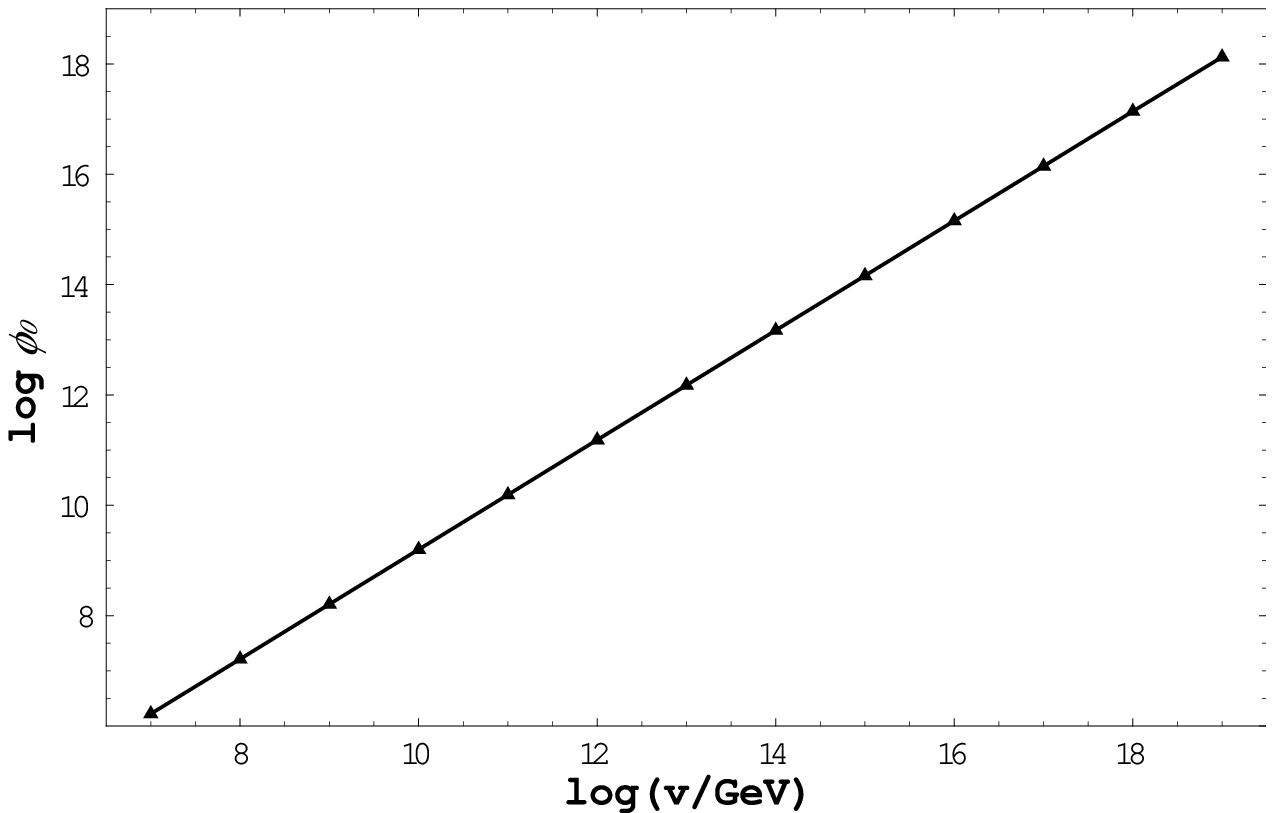}}
 \put (-40,393){(a)}
 \put (80,400){\includegraphics[width=7.5cm,height=5.5cm]{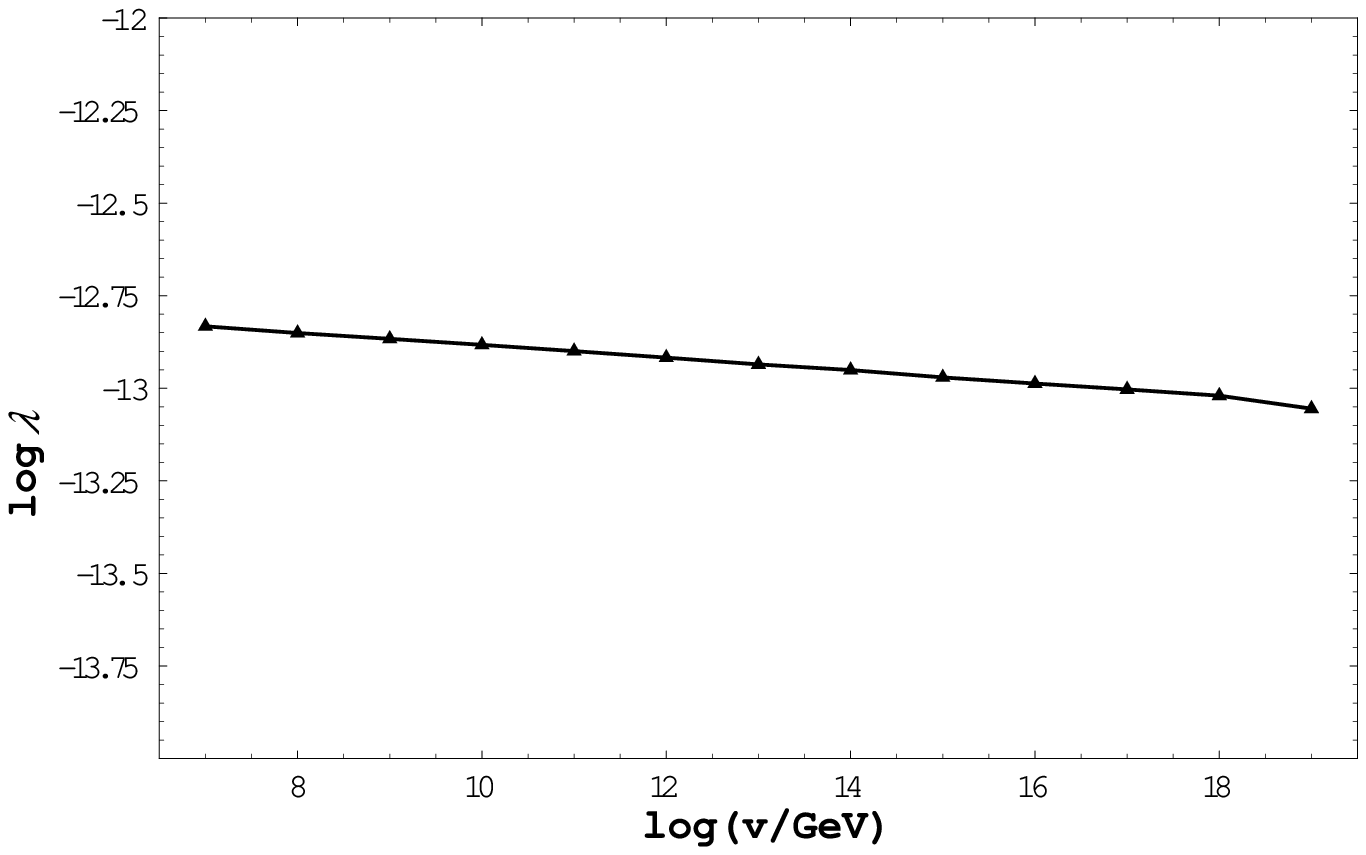}}
 \put (200,393){(b)}
 \put (-150,225){\includegraphics[width=7.5cm,height=5.5cm]{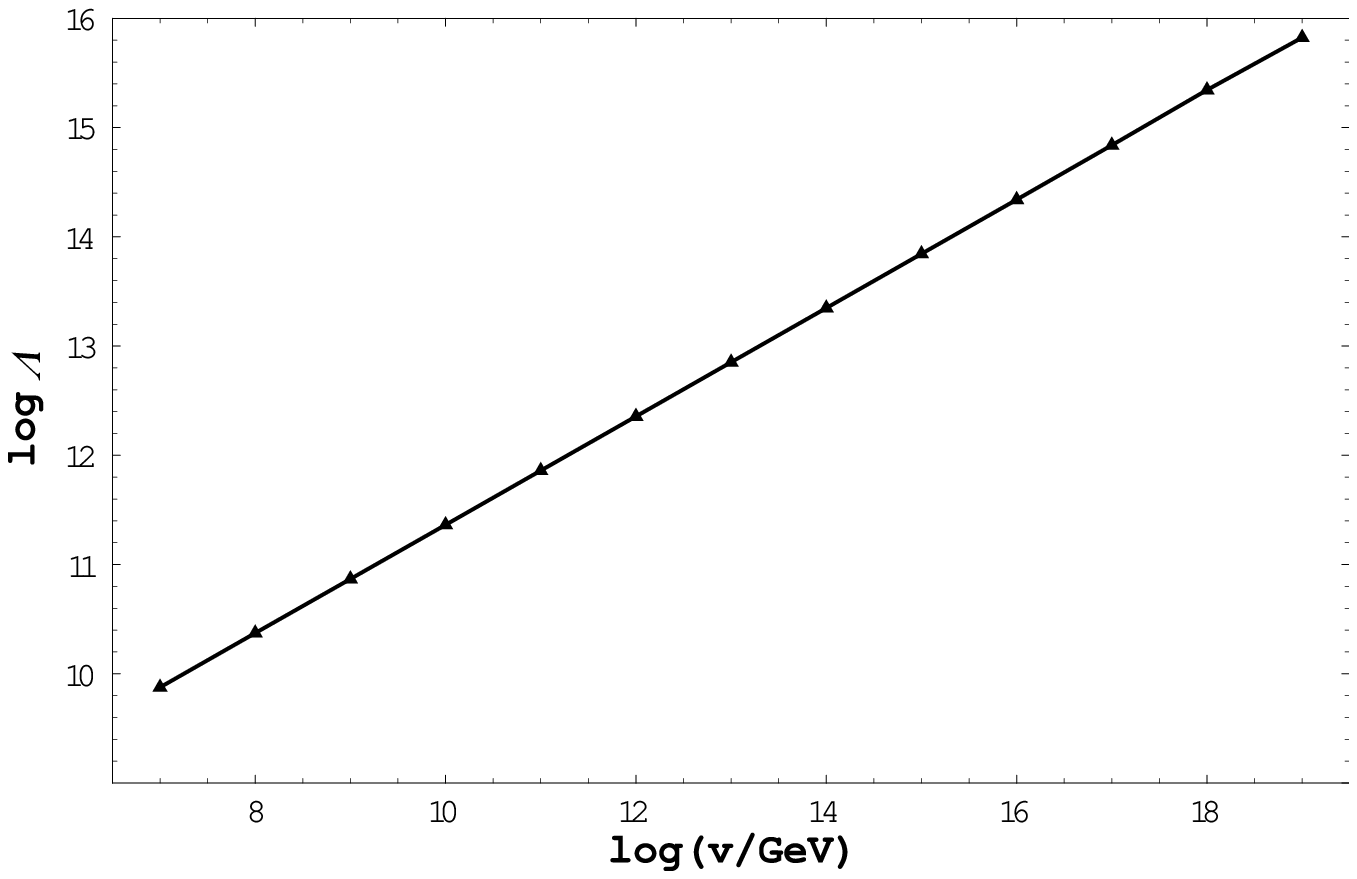}}
 \put (-40,218){(c)}
 \put (80,225){\includegraphics[width=7.5cm,height=5.5cm]{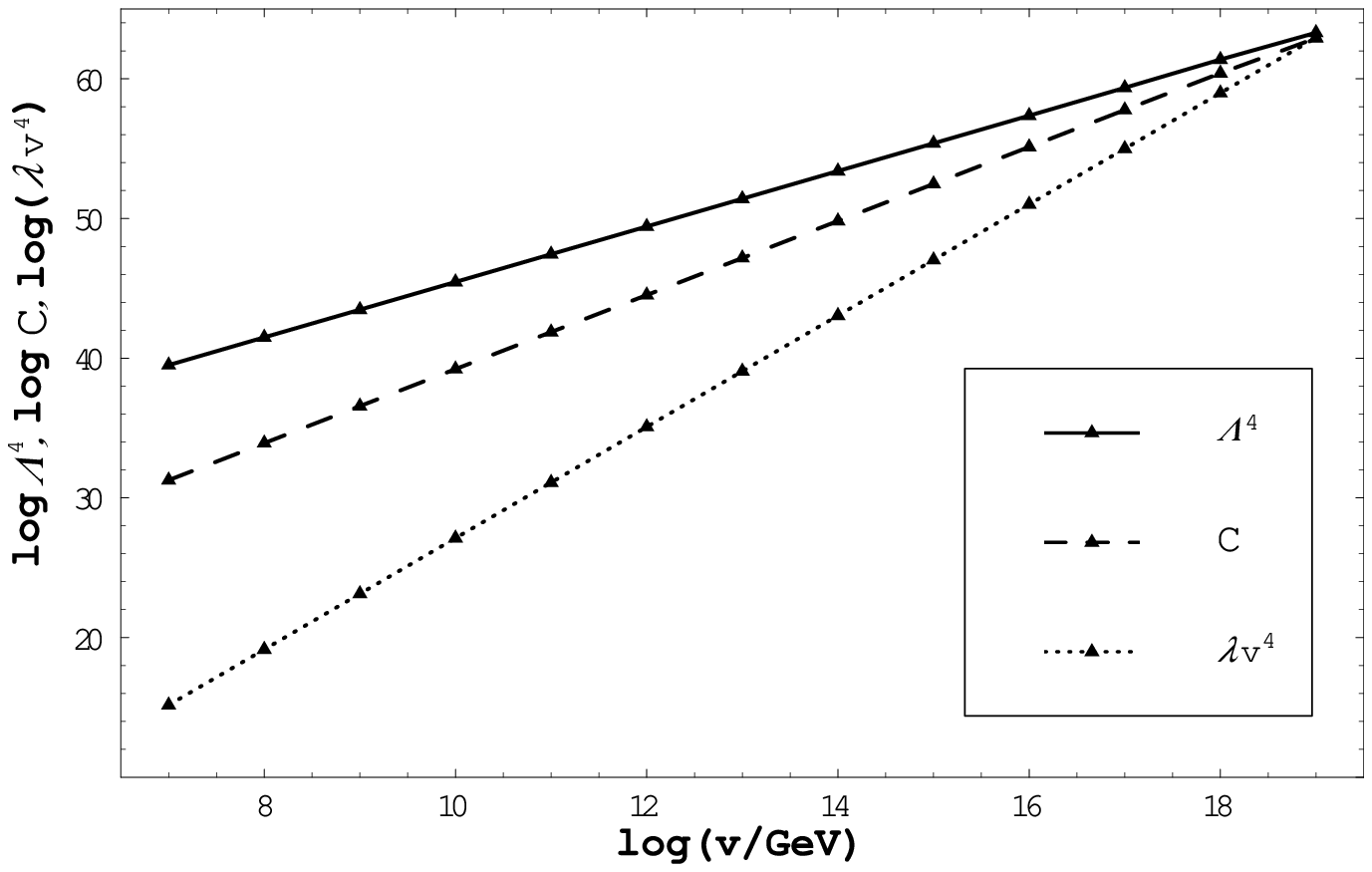}}
 \put (200,218){(d)}
 \put (-150,52){\includegraphics[width=7.5cm,height=5.5cm]{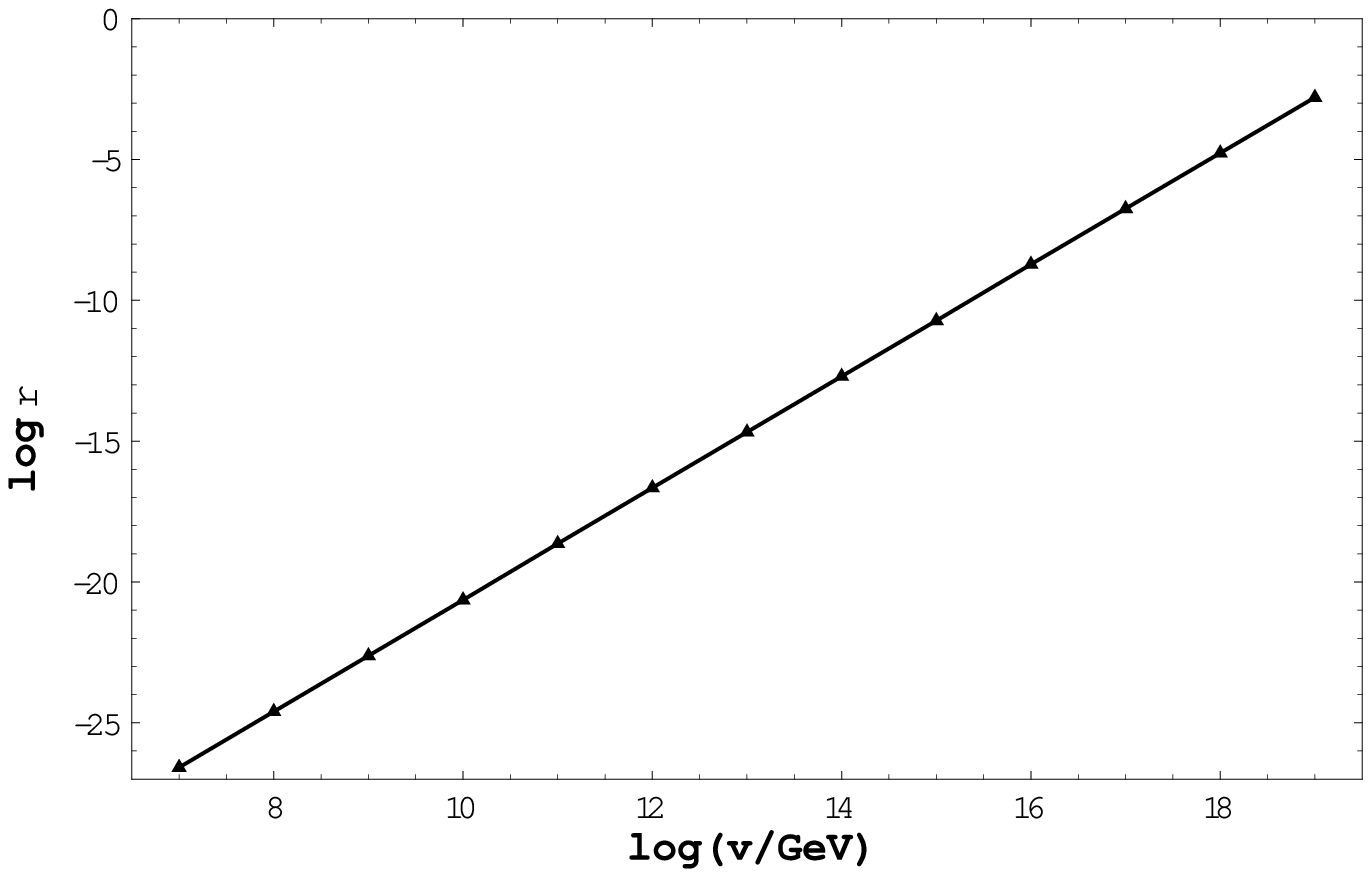}}
 \put (-40,43){(e)}
 \put (80,52){\includegraphics[width=7.5cm,height=5.5cm]{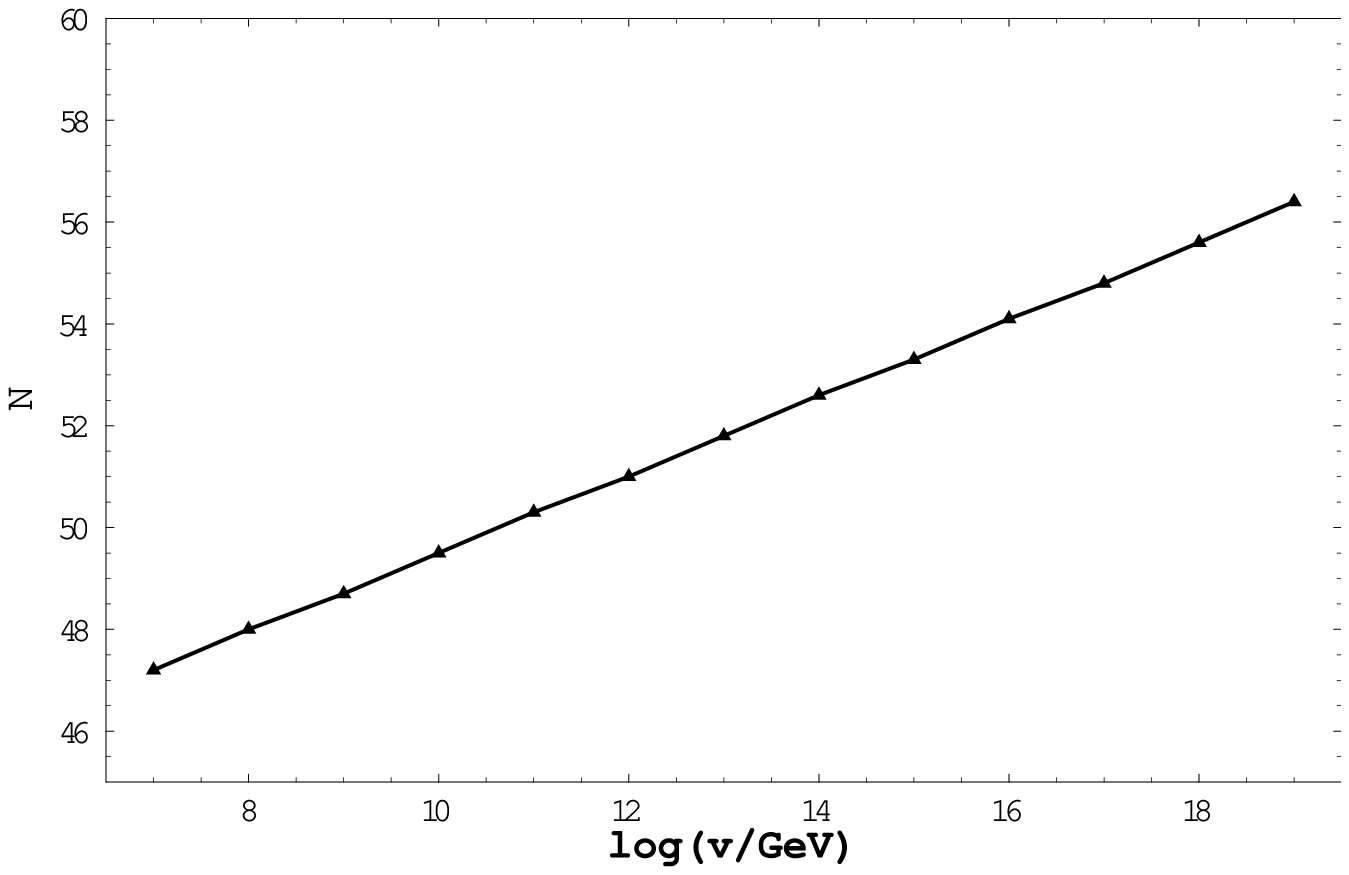}}
 \put (200,43){(f)}
 \end{picture}
\caption{$v-$dependence of various parameters: (a)$-$the inflaton
field value corresponding to the moment when the scale $k_0=0.002
\ {\rm Mpc}^{-1}$ left the horizon, $\phi_0(v)$; (b)$-$the
inflaton self-coupling constant, $\lambda(v)$; (c)$-$the scale of
inflation, $\Lambda(v)$; (d)$-$comparison between $\Lambda^4(v)$,
$C(v)$ and $\lambda v^4$, which tend to be of the same order for
$v\sim M_{\rm P}$; (e)$-$the tensor-to-scalar ratio, $r(v)$;
(f)$-$the number of e-folds of inflation that occur between the
largest observable scale left the horizon and the end of
inflation, $N(v)$.} \label{fig2}
\end{figure}

In Fig.\ref{fig2} we display some graphics with the $v-$dependence
of relevant parameters of our model. In Fig.\ref{fig2}(a) we plot
the numerical results for $\phi_0(v)$, which are then used to
produce the other graphics. From Fig.\ref{fig2}(b), one can see
that $\lambda$ does not vary too much with $v$ and its values are
around $10^{-13}$ for a large range of $v$. We also notice that
the other parameters grow as different powers of $v$. For example
in Fig.\ref{fig2}(c), $\Lambda$, which sets the inflationary
scale, varies as $v^{1/2}$ from $\sim 10^{10}\gev$, for
$v=10^7\gev$, to $\sim 10^{16}\gev$, for $v\sim M_{\rm P}$, i.e.,
from a relatively low-scale to a GUT-scale inflation. In
Fig.\ref{fig2}(d) are represented in the same graphic the values
of $\Lambda^4$, $C$ and $\lambda v^4$ to confirm that, in the
limit $v\ll M_{\rm P}$, one can use the approximation $\lambda
v^4\ll C\ll \Lambda^4$, while for $v\lesssim M_{\rm P}$ the three
terms become of the same order and the above approximation is not
valid anymore. In Fig.\ref{fig2}(e) and (f) we give additional
results, such as the number of "observable" inflation $N$ and the
tensor-to-scalar ratio $r\equiv 16\epsilon$.

In particular, for the lowest possible value $v=10^7\gev$, we get
$N\simeq 47$ e-folds of inflation, and a very tiny value for the
tensor-to-scalar ratio, $r\sim 10^{-27}$, making the detection of
gravitational waves a practically impossible task. We specify that
the spectral index $n_{\rm s}\simeq 0.95$ and the amplitude of
curvature perturbations, ${\cal P_{R}}^{1/2}\simeq 4.86\times
10^{-5}$ for all $v$.

We also notice that for $v\sim M_{\rm P}$, we recover the
inflationary scenario proposed in Paper 2, where $\theta$ was a
quintessence field. The numerical analysis presented here can also
be applied to that model, and one obtains $\langle\phi\rangle \sim
v$, $N\sim 56$ e-folds of inflation and the more interesting
result $r\sim 10^{-3} - 10^{-4}$, which makes gravitational waves
detection more plausible in the future.


\section{Discussions and Conclusions}\label{Conclusions}
In this work, we have presented a model that is able to describe
inflation and dark matter in a unified scenario, by introducing a
new complex scalar field $\Psi=\phi\exp(\rmi\tilde\theta)$
interacting with a real scalar, $\chi$, and a potential invariant
under certain global $U(1)$ symmetry. We allowed for a small
explicit breaking term in the effective potential that is due to
Planck-scale physics and investigated the possibility that $\phi$
is the inflaton and $\theta$ a dark matter particle. The
corresponding constraints have been enumerated in subsections
\ref{Inflation} and \ref{dark_matter}.


In this way, we improve the model of Paper 1, where $\theta$ was a
DM particle, but the model did not include inflation. The results
of Paper 1 are used here in subsection \ref{dark_matter}. For the
part regarding inflation in our model, in subsection
\ref{Inflation} we make a similar analysis as in Paper 2, which
also improves Paper 1 by incorporating inflation, but the
difference is that there $\theta$ was a quintessence field. The
numerical analysis we present in section \ref{Results}
extrapolates between the two scales considered in the models of
Paper 1 and Paper 2.

In the present numerical study, we used the value
$\kappa=10^{-2}$, and we chose $m_{\chi}=\kappa v/2$ for
simplicity. We observe that a tiny value is needed for
$\lambda\sim 10^{-13}$, in order to generate the correct values of
the amplitude of curvature perturbations and of the spectral
index. We have no possible theoretical explanation for justifying
this small $\lambda-$value, but this is a common problem of most
of the inflationary models. Although we make a general numerical
analysis to see how the parameters depend on the SSB scale $v$, we
are finally interested in the value for which the angular field
$\theta$ is a DM candidate, $v\ll M_{\rm P}$.

Notice that for $v\ll M_{\rm P}$, the vev of the inflaton field
$\phi$ is different from $v$ and is approximately given by
$\langle\phi\rangle\simeq v^{2/3}M_{\rm P}^{1/3}\neq v$, while for
$v\sim M_{\rm P}$ they tend to be of the same order, $\langle\phi
\rangle\sim O(v)$.

We included explicit $U(1)-$breaking terms in the potential and
studied the possibility that the resulting PGB, $\theta$, could be
a DM particle. We found in Eq.(\ref{g_cond2}) that the effective
$g$ coupling related to the explicit breaking should be
exponentially suppressed, $g< 10^{-30}$. This confirms our
expectations commented in the Introduction, that the effect of
Planck-scale physics in breaking global symmetries should be
exponentially suppressed \cite{Kallosh:1995hi}. With the extreme
values $g=10^{-30}$ and $v=10^7\gev$, the mass of $\theta$ is
fully determined, $m_{\theta}\sim 15$ eV.

It would be interesting to investigate reheating in our model to
determine the exact reheating temperature $T_{\rm rh}$, and also
to provide a specific mechanism for producing SM particles, but
this goes beyond the scope of our paper.

As a final comment, we would like to add that such a strong
suppression of $g$ may be avoided if, for some reason, $n=7$ and
all smaller values prohibited. In this case, one obtains $g$ of
$O(1)$, but then one should find an argument why $n$ cannot be
smaller than 7.

\appendix
\section{}
\label{Appendix}

Here we give the details of how to obtain the parameters related
to inflation in our model. We base our analysis on 3 constraints,
which are:
\begin{enumerate}
  \item[1.] The number of e-folds of inflation between a given
  scale $k$ crosses the horizon and the end of inflation is given by
     \cite{N_efolds}
      \begin{equation}
         N\simeq 62-\ln\frac{k}{a_0 H_0}-\ln\left(
         \frac{10^{16}\gev}{V_{\rm sym}(\phi_0,0)^{1/4}}\right)+
         \frac13\ln\left(\frac{T_{\rm rh}}{V_{\rm sym}(\phi_0,0)^{1/4}}
         \right)\label{efolds1}
      \end{equation}
  where $T_{\rm rh}$ is the reheating temperature and $a_0/k=H_0^{-1}\simeq 4000$
  Mpc is the biggest observable scale. A scale of interest is $k_0=0.002$
  Mpc$^{-1}$, for which we have reliable observational data \cite{WMAP3}.
  The number of e-folds corresponding to $k_0$ can be expressed
  in terms of the inflaton field $\phi_0$
    \begin{eqnarray}
         N(\phi_0)&=&\frac{8\pi}{M_{\rm P}^2}\int_{\phi_{\rm end}}^{\phi_0}
         \frac{V_{\rm sym}}{V'_{\rm sym}}{\rm d}\phi\nonumber\\[8pt]
         &=&\frac{\pi}{M_{\rm P}^2}
         \left[\frac{4(\Lambda^4-C)}{\lambda v^2}\ln\frac{(v^2-\phi_0^2)\phi_{\rm end}^2}
         {(v^2-\phi_{\rm end}^2)\phi_0^2}+v^2\ln\frac{\phi_{\rm end}^2}{\phi_0^2}-(\phi_{\rm
         end}^2-\phi_0^2)\right].\label{efolds2}
     \end{eqnarray}
     From now on we set $T_{\rm rh}=10^9\gev$ and $\phi_{\rm end}=v/2
     (\equiv m_{\chi}=\kappa v/2)$, for simplicity. By equating the two
     expressions (\ref{efolds1}) and (\ref{efolds2}) for $N(\phi_0)$, we finally obtain
    \begin{eqnarray}
         60&-&\ln
         \frac{10^{16}\gev}{\left[\Lambda^4-C+\frac14\lambda(v^2-\phi_0^2)^2
         \right]^{1/4}} +\frac13\ln\frac{10^9\gev}{\left[\Lambda^4-C+\frac14
         \lambda(v^2-\phi_0^2)^2\right]^{1/4}}\nonumber\\[8pt]
         &=&\frac{\pi}{M_{\rm P}^2}
         \left[\frac{4(\Lambda^4-C)}{\lambda v^2}\ln\frac{v^2-\phi_0^2}
         {3\phi_0^2}+v^2\ln\frac{v^2}{4\phi_0^2}- \left(\frac{v^2}4-
         \phi_0^2\right)\right].
         \label{n_efolds}
     \end{eqnarray}
  \item[2.] The amplitude of the curvature perturbations ${\cal
  P_R}^{1/2}$ has the observed value ${\cal P_R}^{1/2}\simeq 4.86\times
  10^{-5}$ corresponding to the scale $k_0$ \cite{WMAP3}. This means that
     \begin{eqnarray}
         {\cal P_R}^{1/2}&=&\sqrt{\frac{128\pi}3}\left|\frac{V_{\rm sym}(\phi_0,0)^{3/2}}
       {M_{\rm P}^3 V'_{\rm sym}(\phi_0,0)}\right|\nonumber\\[8pt]
       &=&\sqrt{\frac{128\pi}3}
       \frac{\left[\Lambda^4-C+\frac14\lambda(v^2-\phi_0^2)^2\right]^{3/2}}
       {\lambda M_{\rm P}^3 \phi_0(v^2-\phi_0^2)}\simeq 4.86\times 10^{-5}.
         \label{curv_ampl}
      \end{eqnarray}
  \item[3.] The value of the spectral index $n_{\rm s}\simeq 1+2\eta$
  should have a value close to $n_{\rm s}=0.95$ for the same scale $k_0=0.002$
  Mpc$^{-1}$, where $\eta=(M_{\rm P}^2/8\pi)
  (V''_{\rm sym}/V_{\rm sym})$ is a slow-roll parameter. This
  becomes
     \begin{equation}
         2\eta=\frac{-\lambda M_{\rm
         P}^2(v^2-3\phi_0^2)}{4\pi\left[
         \Lambda^4-C+\frac14\lambda(v^2-\phi_0^2)^2\right]}\simeq -0.05.
       \label{spec_ind}
      \end{equation}
\end{enumerate}

One can see that by combining equations (\ref{curv_ampl}) and
(\ref{spec_ind}) one obtains an expression for $\lambda$ in terms
of $\phi_0$ and $v$
\begin{equation}
  \lambda=4.4\times
  10^{-12}\frac{\phi_0^2(v^2-\phi_0^2)^2}{(v^2-3\phi_0^2)^3}.
       \label{lambda}
      \end{equation}
Next, by replacing (\ref{lambda}) into (\ref{spec_ind}) one
obtains
\begin{equation}
  \Lambda^4-C=4.4\times 10^{-12}\frac{\phi_0^2(v^2-\phi_0^2)^2}{(v^2-3\phi_0^2)^2}
  \left[\frac5{\pi}M_{\rm P}^2-\frac{(v^2-\phi_0^2)^2}{4(v^2-3\phi_0^2)}\right].
       \label{lambda-C}
      \end{equation}
Finally, by introducing (\ref{lambda}) and (\ref{lambda-C}) into
(\ref{n_efolds}), one obtains an equation which relates $\phi_0$
and $v$. We solved it numerically for a few $v-$values in the
interval $(10^7-10^{19})\gev$ and we obtained the corresponding
values for $\phi_0$, which are shown in Fig.\ref{fig1}(a). Once we
have $\phi_0(v)$, we can turn back to (\ref{lambda}) and
(\ref{lambda-C}) and find the values of $\lambda$ and
$\Lambda^4-C$, respectively.

Still, we would like to find $\Lambda$ and $C$, separately. This
can be done by requiring that the absolute minimum of $V_{\rm
sym}(\phi,\chi)$ is equal to zero. The position of the absolute
minimum is given by the following conditions
\begin{eqnarray}
  \frac{\partial V_{\rm sym}(\phi,\chi)}{\partial\phi}&=& -\lambda
  v^2\phi+\lambda\phi^3-\kappa^2\phi\chi^2+\frac{\kappa^4\phi^3\chi^4}
  {4\Lambda^4}=0,\;\;\frac{\partial^2 V_{\rm sym}(\phi,\chi)}
  {\partial\phi^2}>0\label{min1}\\[8pt]
  \frac{\partial V_{\rm sym}(\phi,\chi)}{\partial\chi}&=&
  m_{\chi}^2\chi-\kappa^2\phi^2\chi+\frac{\kappa^4\phi^4\chi^3}
  {4\Lambda^4}=0,\;\;\frac{\partial^2 V_{\rm
  sym}(\phi,\chi)}{\partial\chi^2}>0.\label{min2}
       \label{abs_min}
      \end{eqnarray}

Solving the above equation system, one can obtain $\phi_{\rm min}$
and $\chi_{\rm min}$. We do not show here the analytical solutions
because they are very complicated. From the condition $V_{\rm
sym}(\phi_{\rm m},\chi_{\rm m})=0$, one can obtain a relation
between $C$ and $\Lambda$. With this, going back to equation
(\ref{lambda-C}), one obtains the dependence $\Lambda(v)$ and
subsequently $C(v)$. The results we obtained are shown in
Fig.\ref{fig1} (c) and (d), respectively.

Things become much simpler in the limit $v\ll M_{\rm P}$. As shown
in Fig.\ref{fig1} (d), when $v\ll M_{\rm P}$, the following
relations are satisfied
\begin{equation}
  \lambda v^4\ll C \ll \Lambda^4.
\end{equation}
In this case, from (\ref{spec_ind}) we obtain
\begin{equation}
  \Lambda^4\simeq \frac5{\pi} \lambda M_{\rm P}^2(v^2-3\phi_0^2)\sim
  \lambda v^2 M_{\rm P}^2\label{Lambda_approx}
\end{equation}
and the solutions of (\ref{min1}) and (\ref{min2}) become very
simple
\begin{equation}
  \phi_{\rm m}\simeq\frac{v^{1/3}\Lambda^{2/3}}{\lambda^{1/6}}\sim
  v^{2/3}M_{\rm P}^{1/3}\,,\;\;
  \chi_{\rm m}\simeq\frac{2\lambda^{1/6}\Lambda^{4/3}}{\kappa
  v^{1/3}}.
  \label{phi_min}
\end{equation}
With the above expressions, the approximate solution for $C$ is
also very simple
\begin{equation}
  C\simeq\frac34\Lambda^4\left(\frac{\lambda
  v^4}{\Lambda^4}\right)^{1/3}\sim \Lambda^4\left(\frac{v}
  {M_{\rm P}}\right)^{2/3}\sim\lambda v^4\left(\frac{M_{\rm P}}{v}
  \right)^{4/3}
  \label{C_approx}
\end{equation}
where we made use of (\ref{Lambda_approx}). This also helps us
understand why the following inequalities $\lambda v^4\ll C \ll
\Lambda^4$ are satisfied for $v\ll M_{\rm P}$.

In the same limit ($v\ll M_{\rm P}$), we get simple expressions
for the $v-$dependence of the number of e-folds of observable
inflation, $N(v)\propto \ln v$, and of the tensor-to-scalar ratio,
$r(v)\propto v^2/M_{\rm P}^2$.
\begin{acknowledgments}
I thank Eduard Mass{\'o} for useful discussions and comments. This
work is supported by the Spanish grant FPA-2005-05904 and by
Catalan DURSI grants 2005SGR00916 and 2003FI00138.
\end{acknowledgments}


\end{document}